\documentclass[aps,pre,showpacs,amsmath,amssymb,amsfonts,superscriptaddress,lengthcheck,twocolumn]{revtex4-1}
\usepackage{graphicx}
\usepackage{subfigure}
\usepackage{amsthm}
\usepackage{verbatim}
\usepackage{dcolumn}
\usepackage{bm}
\usepackage{epsf}
\usepackage{color}
\usepackage[colorlinks=true,citecolor=blue,linkcolor=blue,urlcolor=blue]{hyperref}%
\usepackage{xcolor}
\usepackage{dsfont}
\newcommand{\bra}[1]{\left\langle #1\right|}
\newcommand{\ket}[1]{\left|#1\right\rangle}

\newcommand{\tr}[1]{\mathrm{tr}\left\{#1\right\}}
\newcommand{\ptr}[2]{\mathrm{tr_{#1}}\left\{#2\right\}}

\newcommand{\e}[1]{\exp{\left(#1\right)}}
\newcommand{\lo}[1]{\ln{\left(#1\right)}}

\newcommand{\com}[2]{\left[#1,\,#2\right]}

\newcommand{\bla}{bla\\bla\\bla\\bla\\bla}

\newcommand{\mc}[1]{\mathcal{#1}}

\newcommand{\mrm}[1]{\mathrm{#1}}

\begin{document}

\title{Quantum to classical transition in an information ratchet}

\author{Josey Stevens}
\affiliation{Department of Physics, University of Maryland Baltimore County, Baltimore, MD 21250, USA}
\affiliation{Applied Physics Laboratory, Johns Hopkins University,  Laurel, MD 20723, USA}
\author{Sebastian Deffner}
\email{deffner@umbc.edu}
\affiliation{Department of Physics, University of Maryland Baltimore County, Baltimore, MD 21250, USA}

\date{\today}

\begin{abstract}
Recent years have seen a flurry of research activity in the study of minimal and autonomous information ratchets. However, the existing classical and quantum models are somewhat hard to compare, and, hence, quantifying possible quantum supremacy in information ratchets has been elusive. We propose a first step towards filling this void between quantum and classical ratchets by introducing a new model with continuous variables -- a quantum particle in a box coupled to a stream of qubits. The dynamics is solved exactly, and we analyze the quantum to classical transition in terms of a natural time scale parameter for the model.
\end{abstract}

\maketitle

\section{Introduction}

It is commonly expected that for certain tasks quantum computers will be exponentially more powerful than classical analogs \cite{Sanders2017}. Loosely speaking this \emph{quantum supremacy} rests in the fact that the quantum logical space is spanned by exponentially more states. For classical computers, Landauer's principle \cite{Landauer1961,Landauer1991} characterizes the minimal, thermodynamic cost necessary to process information. The natural question arises whether this statement of the second law also carries over to quantum systems, or whether quantum effects significantly impact the consumed resources. Most likely the answer to this question will arise from a study of quantum versions of Maxwell's demon \cite{Leff2002}.

Recent years have seen the steady progress in the development of a comprehensive framework for the thermodynamics of information \cite{Sagawa2010,Morikuni2011,DeffnerJarzynski2013,Horowitz2014,Barato2014,Barato2014PRL,Parrondo2015,Strasberg2017}. This has led to the development of minimal classical \cite{Mandal2012,Mandal2013,Barato2013,Boyd2016} and quantum \cite{Zurek1986,Lloyd1997,Quan2006,Kim2011,Deffner2013,Strasberg2013,Park2013,Chapman2015,Elouard2017,Mohammady2017,Franson2017,Zurek2018},  models for information processing, experimental implementations of Maxwell's demon \cite{Koski2015,Vidrighin2016,Camati2016,Cottet2017,Wang2017}, and verifications of Landauer's principle \cite{Berut2012,Jun2014,Hong2016}. 

In the present paper we propose and study a minimal model of a quantum demon \cite{Deffner2013}, aka quantum information ratchet \cite{Sagawa2010}, operating in continuous physical state space.  This analysis is motivated by the classical three state model by Mandal and Jarzynski \cite{Mandal2012} and the discrete, three state quantum model in Ref.~\cite{Deffner2013}.  It is worth emphasizing that these minimal models \cite{Mandal2012,Mandal2013,Deffner2013,Barato2013,Boyd2016,Boyd2016NJP} do not include feedback. In other words, even though information is exchanged with a memory, this information is not ``utilized'' to control the behavior of the system of interest. Rather, these analyses focus on the net effect on the dynamics that arises from the interaction with an information reservoir \cite{DeffnerJarzynski2013}. Here, our main interest lies in how the system behaves as the ratchet transitions from the quantum to classical regime, which permits to directly compare quantum and classical modes of operation. See also Ref.~\cite{Safranek2018} for a study of quantum memories with correlated qubits and the quantum Zeno effect.
\begin{figure}
\includegraphics[width=.48\textwidth]{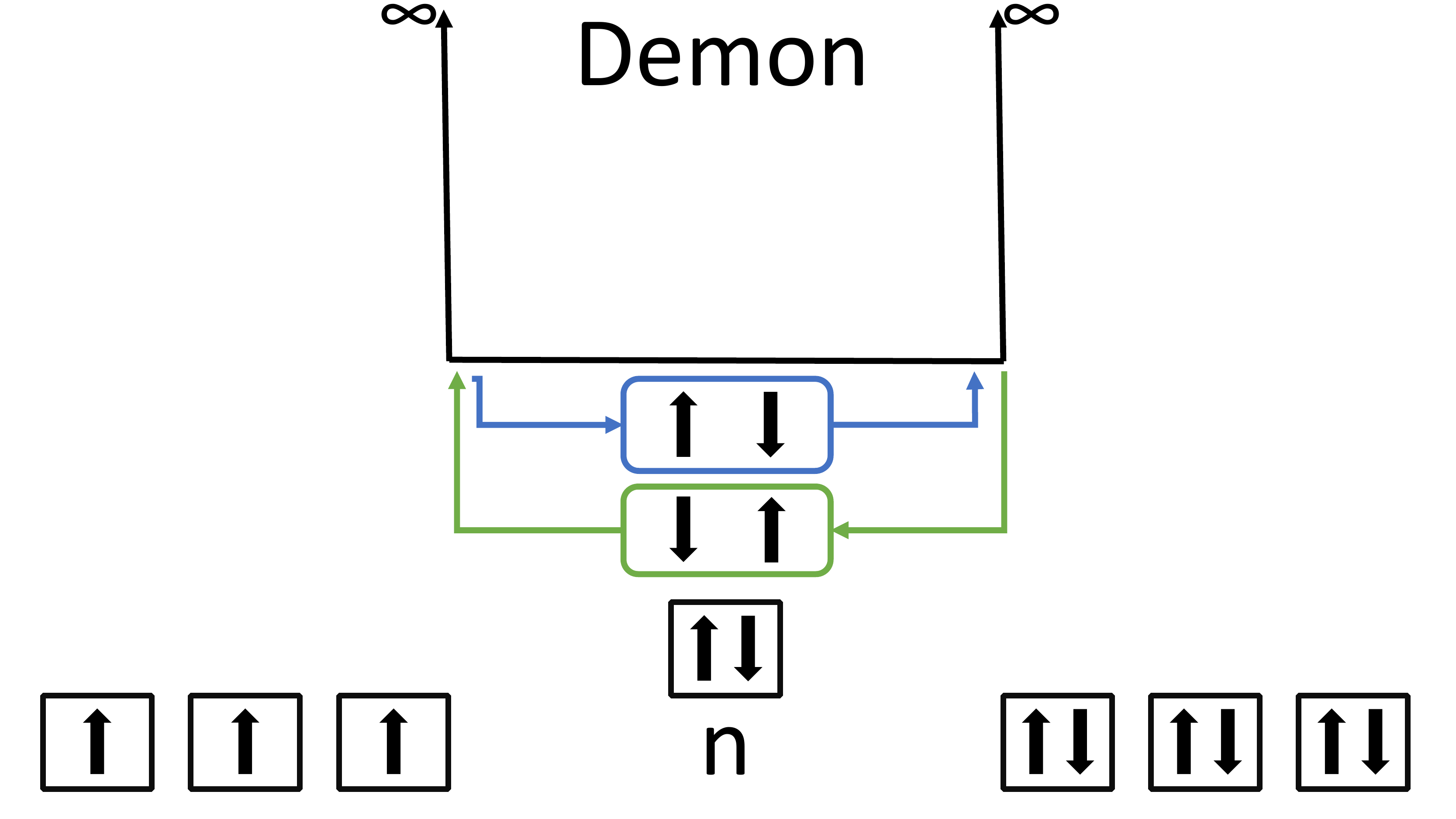}
\caption{\label{fig:Diagram}Schematic illustration of a continuous information ratchet, where the working medium is a quantum particle in a box and the information reservoir is realized as a stream of qubits.} 
\end{figure}

Similarly to the models of Refs.~\cite{Deffner2013} and \cite{Safranek2018}, we analyze the dynamical behavior of a small quantum system, the ``demon'' $\mc{D}$, interacting with a quantum ``memory,'' $\mc{M}$, such that the quantum transitions in $\mc{D}$ become biased. In complete analogy to Refs.~\cite{Deffner2013} and \cite{Safranek2018} the memory $\mc{M}$ is given by a \emph{stream} of identical qubits. For our present purposes, however, $\mc{D}$  is modeled as a particle in a one-dimensional box, which has an infinitely large eigenspectrum. This is in contrast to previous studies \cite{Deffner2013,Safranek2018}, which were limited to single spin-1/2 or spin-1 particles with no clear and systematic classical limit. The dynamics of the ``universe'' spanned by $\mc{D}\otimes\mc{M}$ is assumed to evolve under Schr\"odinger dynamics, which includes qubit coupling, decoupling, and time evolution of the particle in the box.

We solve the dynamics of the continuous quantum ratchet exactly as it writes information into the quantum memory $\mc{M}$. As a main result, we find  that after a transient phase $\mc{D}$ settles into a time periodic steady state with a persistent probability current as information is written into $\mc{M}$. Further introducing a physically motivated time parameter we are able to examine the behavior of the ratchet as it transitions from the deep quantum to the classical regime. Thus, the present analysis constitutes  a solvable, autonomous, and pedagogical example of a quantum demon, or more precisely a quantum information ratchet operating in continuous state space. 

\section{Quantum information ratchet}

The following analysis will study a minimal model of a self-contained quantum information ratchet within the framework of autonomous thermodynamics of information \cite{DeffnerJarzynski2013}. The working medium is a quantum particle, $\mc{D}$, in a one-dimensional box of length $L$ $ ( 0\le x \le L )$ with eigenfunctions and eigenenergies,
\begin{equation}
\label{eq01}
\phi_{n}^L(x)=\sqrt{\frac{2}{L}} \text{sin}\left(\frac{n \pi x}{L}\right)\quad\text{and}\quad E_{n}^L= \frac{(n \hbar \pi)^2}{2 m L ^2}\,,
\end{equation}
where $m$ is the mass of the particle. This particle is coupled to a stream consisting of $N$ qubits, which we denote as $\mc{M}$, see Fig.~\ref{fig:Diagram}.

The dynamics of $\mc{D}\otimes\mc{M}$ is described by the Hamiltonian 
\begin{widetext}
\begin{multline}
\label{eq02}
H_{\mc{D}\otimes\mc{M}} (t)=\\
 \sum_{n=1}^N \sum_{l,i,j} \Pi_n(t)E_{l}^{2L} \ket{\phi_i^L}\bra{\phi_j^L} \left(  a_{il}a^*_{jl}\otimes \mathds{I}_1 \otimes ...\otimes \left(\begin{array}{cc}
1 & 0 \\ 
0 & 0
\end{array}\right)_n 
\otimes...\otimes\mathds{I}_N
+ b_{il}b^*_{jl}
\otimes \mathds{I}_1 \otimes ...\otimes \left(\begin{array}{cc}
0 & 0 \\ 
0 & 1
\end{array}\right)_n 
\otimes...\otimes\mathds{I}_N \right)
\\
 + \sum_{n=1}^N \sum_{l,i,j} \Pi_n(t)E_l^{2L}  \ket{\phi_i^L}\bra{\phi_j^L} \left( a^*_{il}b_{jl} \otimes \mathds{I}_1 \otimes ...\otimes \left(\begin{array}{cc}
0 & 0 \\ 
1 & 0
\end{array}\right)_n 
\otimes...\otimes\mathds{I}_N
+  a_{il}b^*_{jl} \otimes \mathds{I}_1 \otimes ...\otimes \left(\begin{array}{cc}
0 & 1 \\ 
0 & 0
\end{array}\right)_n 
\otimes...\otimes\mathds{I}_N \right)\,,
\end{multline}
\end{widetext}
where $\mathds{I}_n$ is the identity operator in the reduced Hilbert space of the $n$th qubit, $\Pi_n(t)$ is the Heaviside $\pi$ function given by
\begin{equation}
\Pi_n(t) = \begin{cases} 1 &\mbox{if }(n-1)\tau < t \le n \tau \\
0 & \mbox{othewise, }  \end{cases}
\end{equation}
and the coefficients $a_{i,n}$ and $b_{i,n}$ are given by
\begin{equation}
\begin{split}
& a_{i,n}=\left(\bra{\phi_i^L} \otimes \bra{\downarrow}\right)\ket{\phi_n^{2L}} = \int_{-L}^{0}dx\, \phi_n^{2L}(x)\phi_i^L(x+L)\\
& b_{i,n}=\left(\bra{\phi_i^L} \otimes \bra{\uparrow}\right)\ket{\phi_n^{2L}} = \int_{0}^{L}dx\, \phi_n^{2L}(x)\phi_i^L(x)\,.
\end{split}
\end{equation}
Note that in complete analogy to a discrete version of the present model \cite{Deffner2013}, we assume that during the time interval $(n-1)\tau < t \le n \tau$ the ratchet interacts with only the $n$th qubit. 

\begin{figure*}
	\includegraphics[width=.85\textwidth]{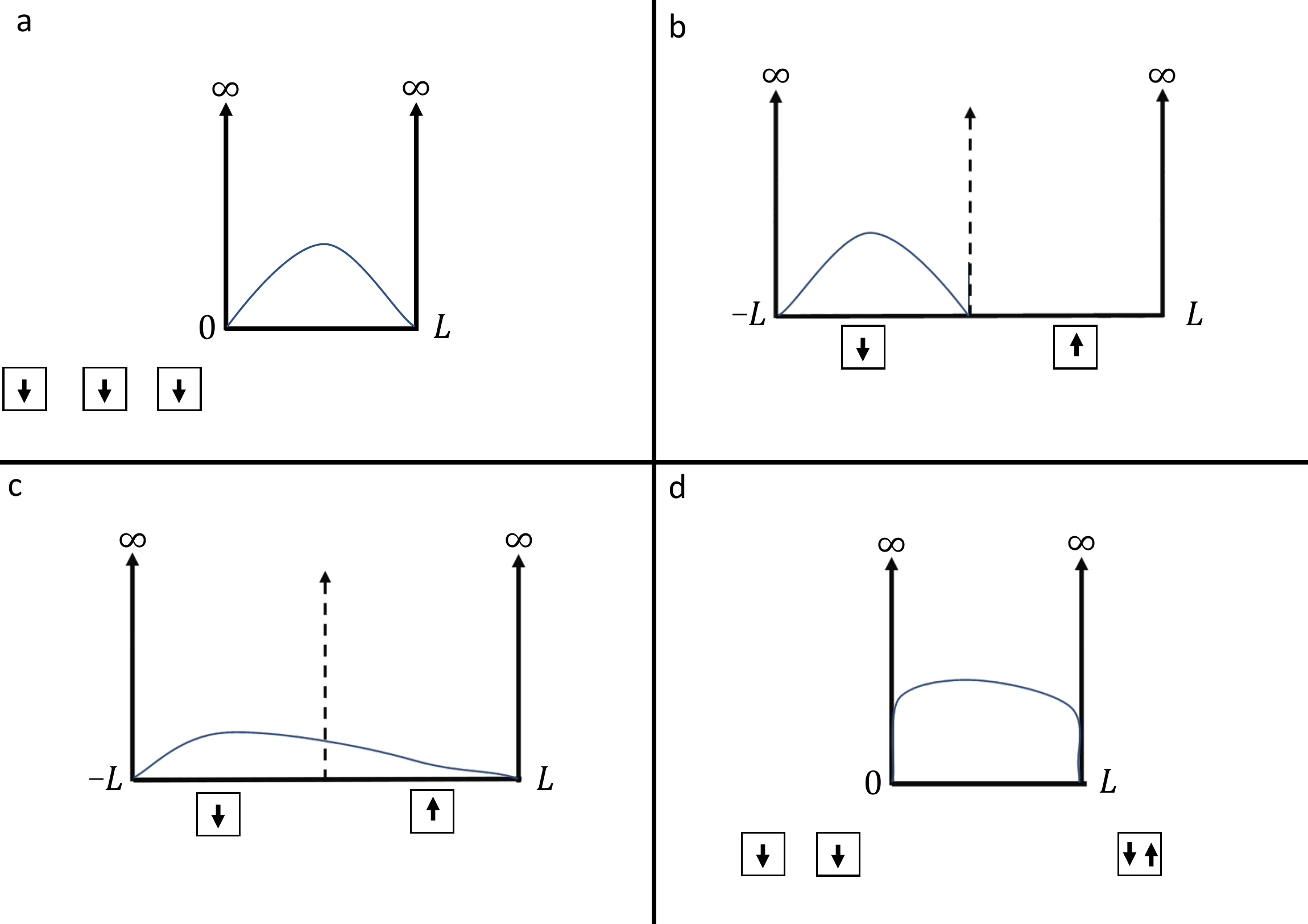}
	\caption{\label{fig:Diagram2} Illustration of the dynamics described by Eq.~\eqref{eq02}: \textbf{a} in the beginning of the $n$th interval $\mc{D}$ is in some state (depicted as solid blue line) and all $(N-n)$ qubits of $\mc{M}$ are in $\ket{\downarrow}$; \textbf{b} same instant as in \textbf{a}, however as represented in the reduced space of $\mc{D}$ and the $n$th qubit of $\mc{M}$; \textbf{c} final quantum state in reduced space of $\mc{D}$ and the $n$th qubit of $\mc{M}$ at the end of the $n$th interval; \textbf{d} same instant as in \textbf{c}, however as represented from the point of view of $\mc{D}$ only; all $(N-n-1)$ qubits in $\mc{M}$ are still in the down state, but the $n$th qubit is now in a superposition of $\ket{\uparrow}$ and $\ket{\downarrow}$.
}
\end{figure*}

Mathematically, the bipartite system composed of $\mc{D}$ and the $n$th qubit can be mapped onto a single particle in a box of with domain $-L \le x \le L$ where the particle occupying $-L \le x \le 0$ corresponds to $\mc{D}$ being in the down state $\ket{\downarrow}$ and the particle occupying $0\le x \le L$ corresponding to the qubit being in the up state $\ket{\uparrow}$. This mapping is shown schematically depicted in Fig.~\ref{fig:Diagram2}.

In the following we will be solving for the reduced dynamics of $\mc{D}$ and the $n$th qubit of $\mc{M}$. Mathematically, we will need to take the partial trace over the $(N-1)$ remaining qubits. To get a little more intuition for the physical dynamics it may be useful to consider that $H_{\mc{D}\otimes\mc{M}} (t)$ is constructed such that when the incoming qubit is in the $\ket{\downarrow}$ state the right wall of the box instantaneously moves expanding the box to length $2L$, i.e., $0\le x \le L \rightarrow 0\le x \le 2L$. Similarly, when the incoming qubit is in the $\ket{\uparrow}$ state the left wall of the box instantaneously moves expanding the box, i.e., $0\le x \le L \rightarrow -L\le x \le L$.
The box instantaneously resets to its original length $L$ when the qubit is decoupled.

This model constitutes a minimal and \emph{autonomous} version of a quantum information ratchet. While our system forgoes coupling to a heat bath and provides no mechanism for work extraction it can still serve as a solvable system to test notions of quantum thermodynamics and the thermodynamics of quantum information processing.

\section{Solution of the dynamics \label{sec:dyn}}
  
The total system, $\mc{D}\otimes\mc{M}$, evolves by the  von-Neumann equation \cite{breuer2002theory} $-i \hbar\, \dot{\rho}_{\mc{D}\otimes\mc{M}} =\com{\rho_{\mc{D}\otimes\mc{M}}}{H_{\mc{D}\otimes\mc{M}}(t)}$, where $\rho_{\mc{D}\otimes\mc{M}}$ is the density operator. The reduced density operator of $\mc{D}$ is obtained by taking the partial trace \cite{Nielsen2010} over $\mc{M}$,
\begin{equation}
\label{partialtrace}
\rho_\mc{D}(t) =\ptr{\mc{M}}{\rho_{\mc{D}\otimes\mc{M}}(t)}=\ptr{\mc{M}}{U(t)\rho_{\mc{D}\otimes\mc{M}} U^\dagger (t)}.
\end{equation}
where $U(t)= \mc{T}_> \e{-i/\hbar \int_{0}^{t}ds\, H_{\mc{D}\otimes\mc{M}}(s)}$

To analyze the dynamics of the ratchet we need to solve for the completely positive trace preserving (CPTP) \citep{Nielsen2010} map which determines the evolution of $\rho(t)$. We can write in Kraus operator expansion   \cite{Nielsen2010},
\begin{equation}
\rho_\mc{D}(t)=\sum_{i}T_i \rho_\mc{D}(t_0) T^\dagger_i\,,
\end{equation}
In general determining $T_i$ for a system consisting of $N+1$ (each qubit plus $\mc{D}$) individual systems is a formidable task. Here the situation is greatly simplified since at time $(n-1)\tau < t \le n\tau$ $\mc{D}$ only interacts with the $n\text{th}$ qubit and the qubits in $\mc{M}$ are independent. For correlated quibits we refer to Ref.~\cite{Safranek2018}. Therefore, the total dynamics can be determined by successively solving the dynamics in the reduced (but still infinite dimensional) Hilbert space of the demon and the $n\text{th}$ qubit. The CPTP map can be constructed by a recursive protocol constructed of Krauss operators $T_i$.

\paragraph*{Step 1:}

At time $t=(n-1)\tau$ the demon is decoupled from the $(n-1)$th qubit, which we denote by $\mc{Q}^{(n-1)}$. Thus, we have
\begin{equation}
\label{eq:ptrace}
\begin{split}
&\rho_\mc{D}(\tau(n-1))=\ptr{\mc{Q}^{(n-1)}}{\rho_{\mc{D}\otimes\mc{Q}^{(n-1)}}(\tau(n-1))}\\
&=P_1 \rho_{\mc{D}\otimes\mc{Q}^{(n-1)}} P_1^\dagger + P_2 \rho_{\mc{D}\otimes\mc{Q}^{(n-1)}} P_2^\dagger\,,
\end{split}
\end{equation}
which we expand in terms of the two Kraus operators $P_1$ and $P_2$. In the reduced Hilbert space of the bipartite state spanned by $\ket{\phi_n}\otimes \mathds{I}_d$, $P_1=\bra{\downarrow} \otimes  \mathds{I}_d$ and $P_2=\bra{\uparrow} \otimes  \mathds{I}_d$ are projection operators into either the left or right side of the box and where $ \mathds{I}_d$ is the identity operator in the Hilbert space of the demon. Further details can be found in Appendix \ref{Appendix}.

\paragraph*{Step 2:}

Immediately after the $(n-1)$th qubit is decoupled from $\mc{D}$ the $n$th qubit is coupled,
\begin{equation}\label{coupling}
\begin{split}
&\rho_{\mc{D}\otimes \mc{Q}^{n}}(\tau(n-1))= \rho_\mc{D}(\tau(n-1)) \otimes \rho_{\mc{Q}^{n}}\\
&=B\rho_{\mc{D}}((n-1)\tau)B^\dagger
\end{split}
\end{equation}
where, in the reduced Hilbert space of the bipartite state spanned by $\ket{\phi_n}\otimes \mathds{I}_d$ and for pure initial states, $B=\mathds{I}_d \otimes \ket{i} $ and $\ket{i}$ is the initial state of the $n\text{th}$ qubit. Again, further details can be found in  Appendix \ref{Appendix}.  It is worth noting that Eq.~\eqref{coupling} would not hold if the demon interacted with multiple qubits at a single instance in time due to quantum correlations and entanglement as this is the case in Ref.~\cite{Safranek2018}.  

\paragraph*{Step 3:}

For $(n-1)\tau < t \le n \tau$ both $\mc{D}$ and $n$th qubit evolve under the unitary dynamics generated by reduced Hamiltonian of $\mc{D}$ and $n$th qubit
\begin{multline}
H_{i,j}= \sum_{l} E_n^{2L}\left(a_{i,l}a^*_{j,l} \otimes \ket{\downarrow}\bra{\downarrow}+b_{i,l}b^*_{j,l}\otimes \ket{\uparrow}\bra{\uparrow}\right)\\ + \sum_{l} E_n^{2L}\left(a^*_{i,l}b_{j,l}\otimes \ket{\downarrow}\bra{\uparrow} +a_{i,l}b^*_{j,l}\otimes \ket{\uparrow}\bra{\downarrow}\right)\,.
\end{multline}
However, since this system is a particle in a one dimensional box of length $2L$, the Hamiltonian can be written in the eigenbasis of the bipartite Hilbert space, $ H_{i,j} = E_i^{2L}\delta_{i,j}$.

\paragraph*{Recursive Map.}

Now that we have the Kraus operators governing each cyclic step we can, in complete analog to Ref.~\cite{Deffner2013}, write down the recursive generator of our CPTP map
 \begin{equation}
 \label{eq:CPTPMap}
 \begin{split}
 &\rho_\mc{D}(n\tau)=U(\tau) B P_1\, \rho_{\mc{D}\otimes\mc{Q}^{(n-1)}}(\tau(n-1))\, P_1^\dagger B^\dagger U^\dagger(\tau)\\
 &\quad + U(\tau) B P_2\, \rho_{\mc{D}\otimes\mc{Q}^{(n-1)}}(\tau(n-1))\, P_2^\dagger B^\dagger U^\dagger(\tau)\,,
 \end{split}
 \end{equation}
 which describes the exact dynamics of the quantum information ratchet.
 
To continue we note that all CPTP maps have a fixed point and that if this fixed point is unique the generated time evolution will converge on this fixed point \cite{Deffner2013,Terhal2000}. For the present dynamics this means that after an initial transient  the dynamics relaxes toward this fixed point. More precisely if $n$ is sufficiently large, then for each time during the cycle $(n-1)\tau < t \le n \tau$ a dynamical fixed point is established. Thus, the demon relaxes into a time-periodic, or stroboscopic steady state where $\rho_\mc{D}(n\tau) \xrightarrow{n \rightarrow \infty}\rho_\mc{D}^\mrm{SS}$. A proof of this statement can be found in in Appendix \ref{AppendixB}.

\paragraph*{Initial preparation of $\mc{M}$.}

In the following we will numerically illustrate the behavior of the continuous, quantum information ratchet for several cases. For the sake of simplicity, we will assume that all $N$ qubits in $\mc{M}$ have been identically and independently prepared in the same state, $\rho_n(0)$. We consider,
\begin{equation}
\label{eq:prep}
\begin{split}
&\rho_n^1(0) = \ket{\downarrow}\bra{\downarrow}\\
&\rho_n^2(0) = \ket{\uparrow}\bra{\uparrow}\\
&\rho_n^3(0) = \left( \ket{\downarrow}\bra{\downarrow}+\ket{\uparrow}\bra{\uparrow}\right)/2\\
&\rho_n^4(0) = \left(\ket{\downarrow}+i\ket{\uparrow} \right)\left( \bra{\downarrow} - i\ket{\uparrow} \right)/2
\end{split}
\end{equation}
States  $\rho_n^1(0)$ and $\rho_n^2(0)$ correspond to a stream of qubits, which are all prepared in the down or up states respectively.
Classically this would be a completely empty memory. State  $\rho_n^3(0)$ represents a stream of qubits where each qubit is equally likely to be in the up or down state, which corresponds to a classically completely full memory (see Appendix \ref{app:zero}). Finally, state $\rho_n^4(0)$ is a deeply quantum qubit stream with no immediate classical analogue. 

To further exclude any predetermined bias in the dynamics of the ratchet, $\mc{D}$ is initially, at $t=0$ prepared in the ground state of the bipartite system. Note, that  the ground state is parity even about the center of the box.

\section{Thermodynamics of information}

\begin{figure}
\includegraphics[width=.48\textwidth]{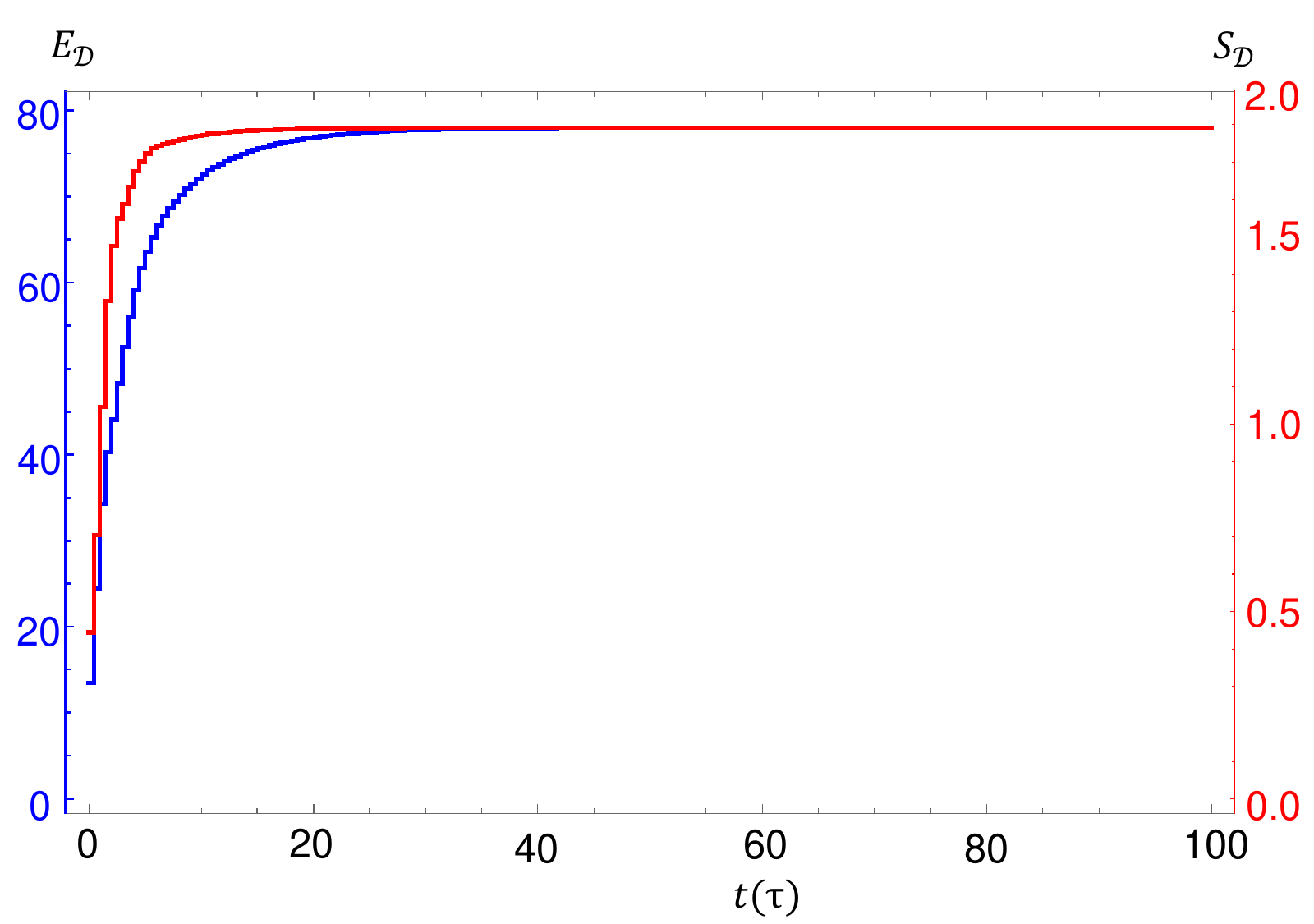}
\caption{\label{fig:Energy}(color online) Energy (lower, blue line) and von Neumann entropy (upper, red line) of the demon as it undergoes repeated qubit interactions.  We see that both increase in discrete steps at each new interaction with a qubit in state $\rho_n^1(0)$ \eqref{eq:prep}.
}
\end{figure}

From the full time evolution of $\rho_{\mc{D}\otimes\mc{M}}$ we can obtain insight into the thermodynamic properties of the information ratchet.   Naturally, prime attention lies on the von Neumann entropy of $\mc{D}$, $S_\mc{D}(t)=-\tr{\rho_\mc{D}(t) \lo{\rho_\mc{D}(t)}}$. 

In Fig.~\ref{fig:Energy} we plot $S_\mc{D}(t)$ together with the average, reduced Hamiltonian, $E_\mc{D}(t)=\tr{\rho_\mc{D}(t) H_\mc{D}(t)}$. We observe that both quantities are monotonically rising until they asymptotically approach their values in the periodic stationary state. It is worth highlighting that in contrast to previous discrete models \cite{Deffner2013}, $\mc{M}$ has an energetic contribution to the dynamics. However, also in the present case $\mc{M}$ does constitute a true information reservoir, as in the stationary state the no energy is exchanged over one cycle of operation. This is precisely in line with the characteristics and definition of information reservoirs \cite{DeffnerJarzynski2013}.


\section{Information driven current}

We concluded the analysis by computing the probability current through the ratchet. In Ref.~\cite{Deffner2013} it was shown that as $\mc{D}$ writes information to the qubit stream the state of $\mc{D}$ undergoes cyclic flow of state occupation probabilities, or a discrete state space probability current. Note that due to the absence of heat and work reservoirs the present ratchets fails ``to do anything useful''. The only analyzable feature of the interaction with $\mc{M}$ is the resulting probability current.

This probability current can be expressed as
\begin{equation}
	j(x,t)= \frac{i \hbar}{2m} [\partial_y \bra{x}\rho(t)\ket{y}-\partial_x \bra{x}\rho(t)\ket{y}]|_{x=y}
\end{equation}
By integrating over the length of the box we can obtain the total probability current of the particle at time $t$
\begin{equation}
\Phi(t)= \int_{-L}^{L} j(x,t) dx
\end{equation}
\begin{figure} 
	\includegraphics[width=.48\textwidth]{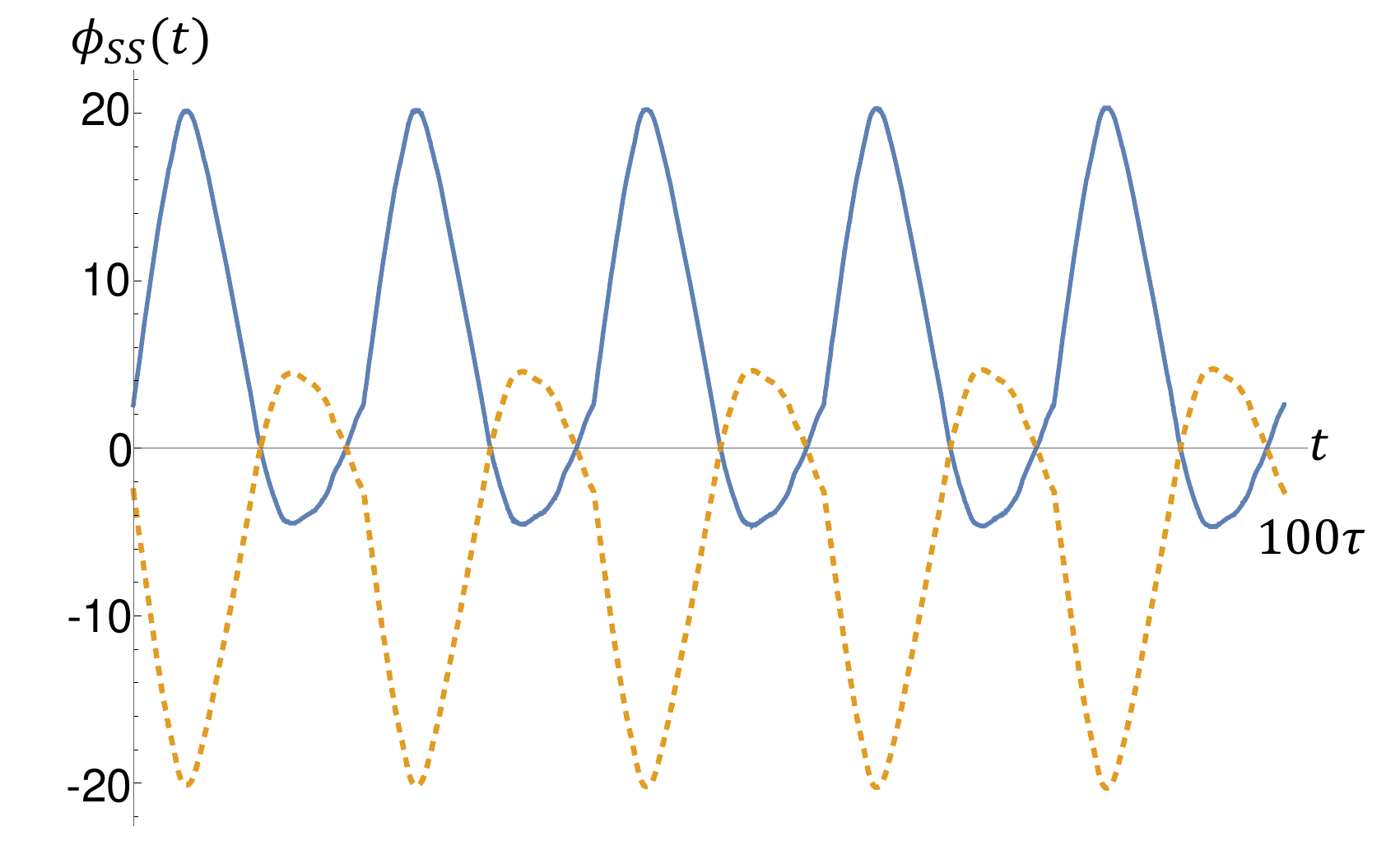}
	\caption{\label{bp1ap2}(color online) Total current $\Phi_{ss}(t)$ in the time periodic steady state with $\tau=\pi \tau_s/30$ for initial qubit preparations of $\rho_n^1(0)$ (blue, solid line) and $\rho_n^2(0)$ (orange, dashed line).
	Only the last five periods are shown. 	
}
\end{figure}
Figure~\ref{bp1ap2} depicts $\Phi(t)$ after $\mc{D}$ has reached its periodic, stationary state. We observe that similarly to the discrete case $\Phi_{ss}(t)$ is an oscillatory function with period of the interaction time $\tau$. The natural question arises, how this probability current behaves as we vary the parameters of our system.

\paragraph*{Parameterizing the Demon.}

To this end, we also compute the average steady state current over one qubit interaction
\begin{equation}
\bar{\Phi}=\frac{1}{\tau} \int_{0}^{\tau}dt \Phi_{SS}(t)\,.
\end{equation}
We immediately see that $\bar{\Phi}$ is dependent on the interaction time $\tau$ but also recall that the dynamics overall depend on the length of the box $L$ and the mass of the particle $m$.

Thus, we introduce the characteristic time parameter, $\tau_s$, which dictates the rate at which our demon evolves in time
\begin{equation}
\tau_s \equiv \frac{8 m L^2}{\pi \hbar}
\end{equation}   
Remarkably, the eigenenergies of the dual demon qubit basis can be written as $E_n^{2L} = \pi \hbar n^2/\tau_s$. Therefore, $\tau_s$ scales the rate at which the each eigenstate of the dual basis evolves in time depending on the particle mass and box size. In this way, $\tau_s$  also quantifies the ``classicallity'' of the particle in a box.

Consider the \emph{classical} limit  $\hbar \rightarrow 0$ and $m \gg 1$. In this case,  $\tau_s$  diverges and the time evolution operator becomes unity
\begin{equation}
\begin{split}
&U_n(t)= \e{-i\frac{n^2 \pi^2 \hbar}{8 m L^2}t} \\
& \ \quad \quad = \e{- i \pi n^2 \frac{t}{\tau_s}} \xrightarrow{\tau_s \rightarrow \infty} \mathds{I}
\end{split}
\end{equation}
This is in full agreement with classical intuition. Namely, in the case of a classical particle in a box, i.e. expanding instantaneously expanding the walls of a box containing an initially stationary classical particle will \emph{not} induce a current in this particle.

\begin{figure} 
	\includegraphics[width=.48\textwidth]{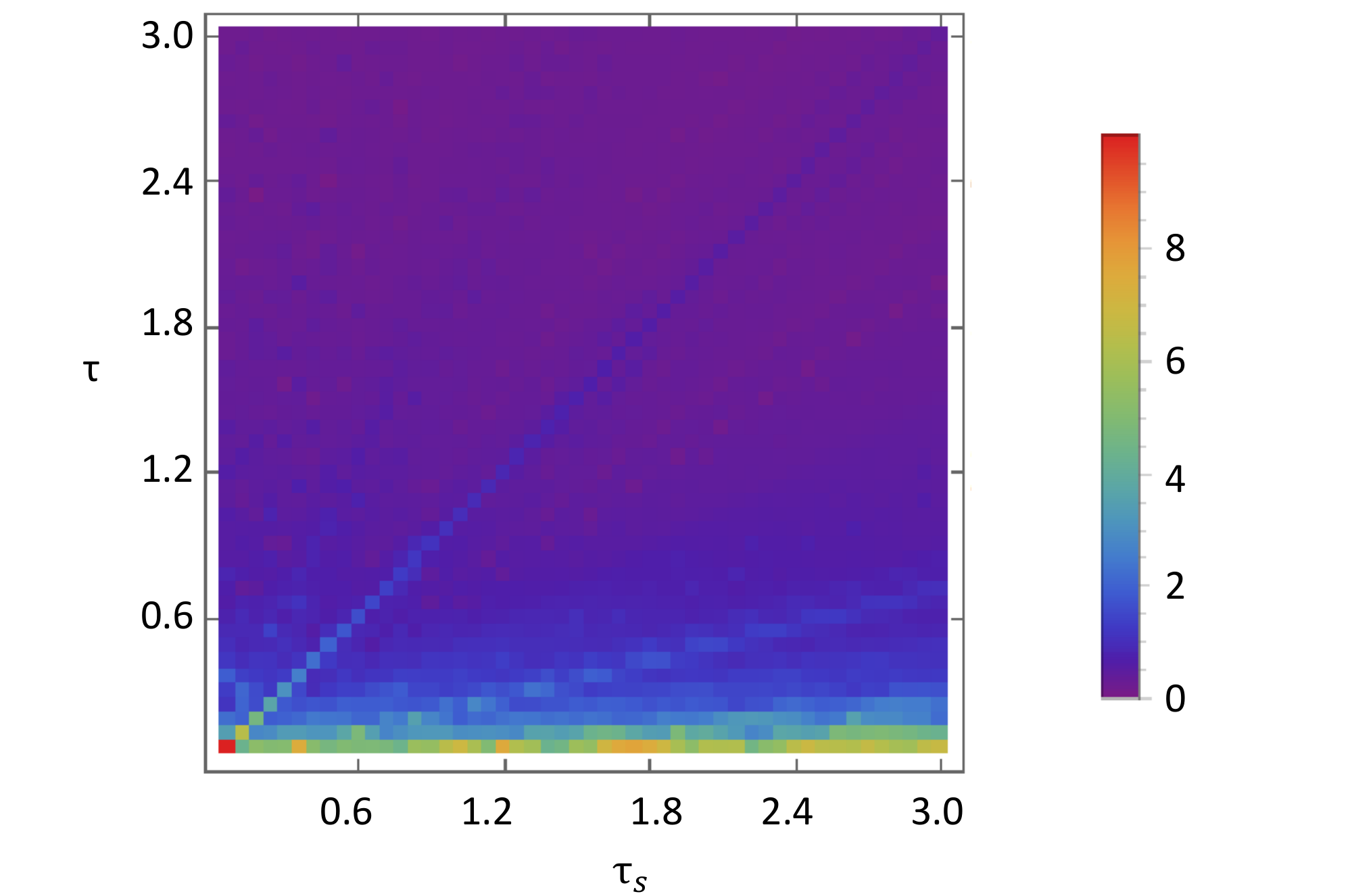}
	\caption{\label{p1}(color online) Total current $\bar{\Phi}$ for $\rho_n(0) = \rho_n^1(0)$ as a function of qubit interaction time $\tau$ and the characteristic time $\tau_s$ where both $\tau$ and $\tau_s$ are given in arbitrary units.  	
}
\end{figure}

\begin{figure}
	\includegraphics[width=.48\textwidth]{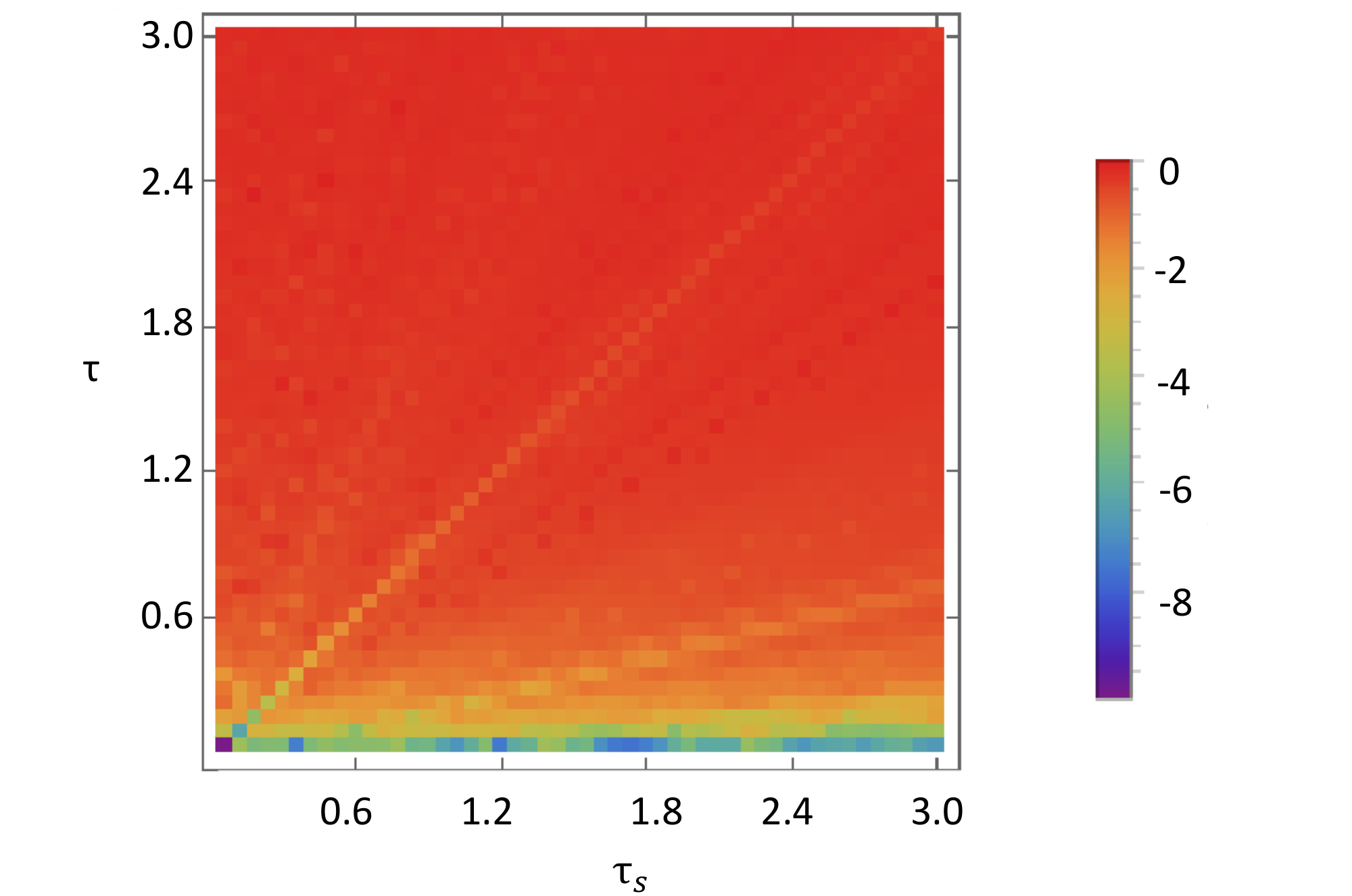}
	\caption{\label{p2}(color online) Total current $\bar{\Phi}$ for $\rho_n(0) = \rho_n^2(0)$ as a function of qubit interaction time $\tau$ and the characteristic time $\tau_s$ where both $\tau$ and $\tau_s$ are given in arbitrary units.  	
}
\end{figure}

\begin{figure}
	\includegraphics[width=.48\textwidth]{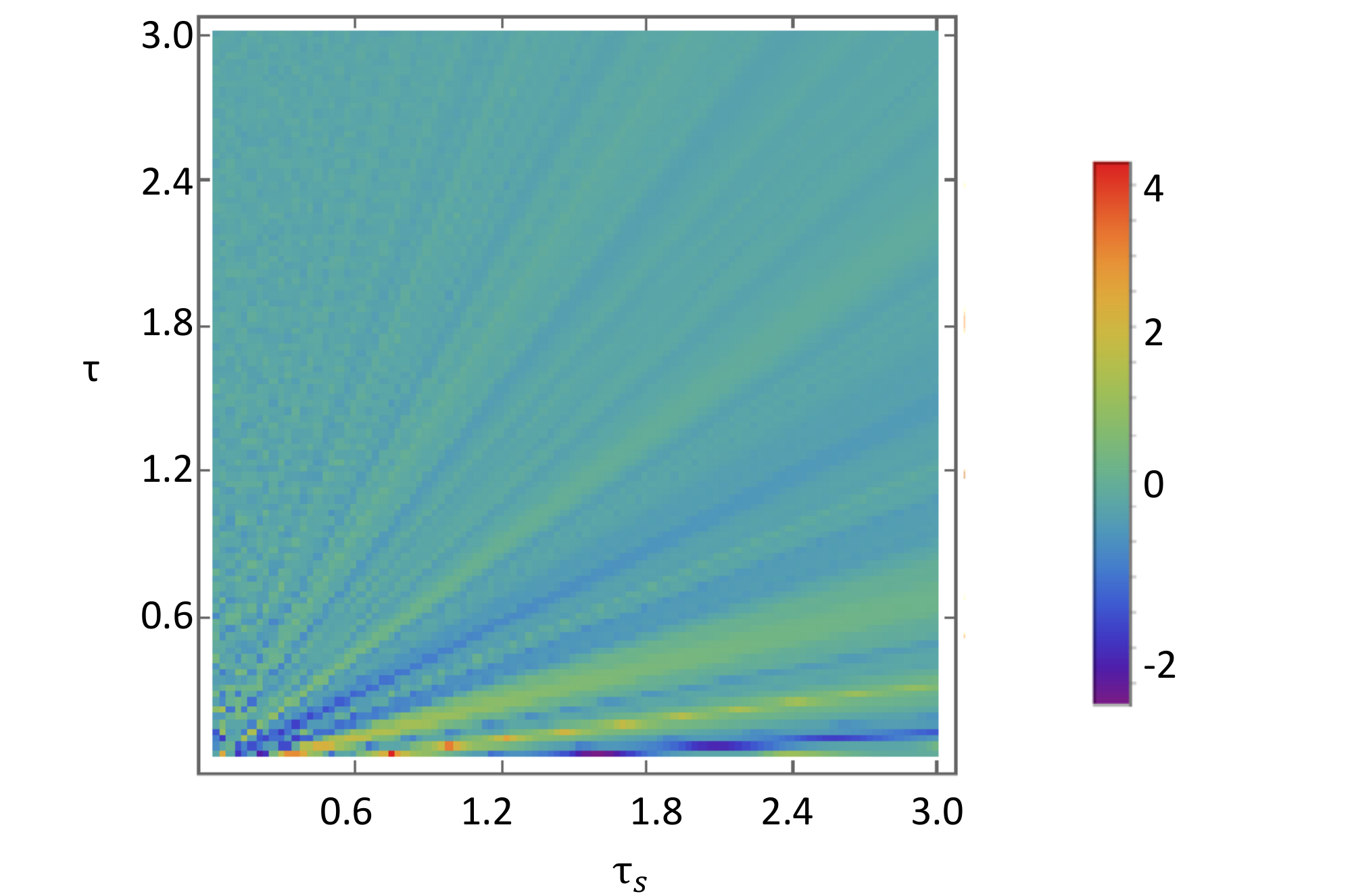}
	\caption{\label{p4}(color online) Total current $\bar{\Phi}$ for $\rho_n(0) = \rho_n^4(0)$ as a function of qubit interaction time $\tau$ and the characteristic time $\tau_s$ where both $\tau$ and $\tau_s$ are given in arbitrary units.  	
}
\end{figure}

In Figs.~\ref{p1}-\ref{p4} we plot the total steady state currents as a function of $\tau$ and $\tau_s$ for each of our qubit preparation states $\rho_n^1(0),\rho_n^2(0),\rho_n^4(0)$ respectively. We see that for $\rho_n^1(0)$ the qubit interaction always induces a positive current in $\mc{D}$. Further, we observe that in order to obtain the largest probability current for any given set of parameters is to drive the system such that the interaction time is equal to the characteristic time, i.e, $\tau = \tau_s$.

Comparing Figs.~\ref{p1} and \ref{p2} we notice that by inverting the initial state of the $N$ qubits the induced current switches sign and is also consistent with the total currents shown in Fig.~\ref{bp1ap2}. This is again in full agreement with physical intuition, namely that the qubit states correspond to the effective interaction shown in FIG. \ref{fig:Diagram}, i.e., expanding the box to either the left or the right. As a final consistency check, we also confirmed that preparing the qubits in the maximum classical information state, $ \rho_n^3(0)$, there is never an induced current since the particle is always equally likely to evolve in either direction (see Appendix \ref{app:zero}).

\paragraph*{Completely quantum current.}

Finally, examining  $ \rho_n^4(0)$, the ``truly" quantum preparation state which has no classical analogue, we observe that the sign is dependent on both $\tau$ and $\tau_s$. This dependence is a purely quantum feature that is caused by quantum correlations of the initial preparation and is the continuous version the interaction strength dependent state current calculated in Ref.~\cite{Deffner2013}. Interestingly, the average, persistent current survives deep into the classical regime. However, due to the deeply quantum nature of $ \rho_n^4(0)$  we would not expect this behavior to appear from any classical preparation of $\mc{M}$, or have a simply and intuitive explanation. 

\paragraph*{Behavior for large $\tau_s$.}

As a final check, we observe that for $\tau_s \gg \tau$ for any fixed driving time the induced current does in fact decrease toward zero.
as in the classical limit $\tau_s \rightarrow \infty$ the total current has to vanish. Therefore, we can conclude that the  induced probability current is an exclusively quantum property of the quantum information ratchet.

\section{Possible extensions of the model}

In the present analysis we have restricted ourselves to the simplest case of a particle in a box coupled to an information reservoir comprised of a stream of qubits.  We have thus forgone the typical treatment of Maxwell's demon or Szilard's engine in which the demon is coupled to a thermal reservoir, whose energy is harnessed by the demon to do work. Indeed the present analysis serves as only a starting point for understanding how the thermodynamics of quantum information differ from those of classical information. 

Here, we have shown that writing quantum information can be used to induce persistent currents. While this current is associated with a directed form of energy we have not demonstrated how to extract work from the ratchet or how the system behaves if coupled to a thermal environment.
Certainly the next step would be to include thermal reservoirs and devise a method for work extraction to answer questions more closely related to thermodynamics.

\section{Concluding Remarks}

In summary we have proposed and analyzed a simple, solvable, and pedagogical example of an autonomous quantum information ratchet, which operates in a continuous state space. We have demonstrated that as information is written into a stream of qubits a persistent steady state current is induced, which is consistent with previously analyzed models of quantum and classical Maxwell's demon. The continuous spectrum has made it possible to introduce a simple measure of classicallity, and we have analyzed the behavior of the persistent current in the classical limit. As main insight, we have concluded that that in comparing quantum and classical models for information processing differences in behavior may not be due solely on information type, but also in the very nature of quantum and classical dynamics. In addition, we have shown that truly quantum information states unlock modes of operation which persist in both quantum and classical regimes.

\acknowledgements{We would like to thank Cleverson Cherubim for fruitful and enjoyable discussions. J.S. acknowledges a GAANN Fellowship from the Department of Education (P200A150003). S.D. acknowledges support from the U.S. National Science Foundation under Grant No. CHE-1648973.}

\appendix

\section{Determining the Kraus Operators}
\label{Appendix}

In Sec.~\ref{sec:dyn} we solve the dynamics of the bipartite demon-qubit system.
In this analysis we introduced the projection operators $P_1=\bra{\downarrow} \otimes  \mathds{I}_d$ and $P_2=\bra{\uparrow}   \otimes  \mathds{I}_d$ used in  Eq.~\eqref{eq:ptrace} to define the partial trace $B=\mathds{I}_d \otimes \ket{i}$.

We can explicitly calculate our partial trace over the information bearing qubits as
\begin{equation}
\begin{split}
&\rho_\mc{D} = \ptr{\mc{Q}}{\rho}\\
&= (\bra{\downarrow}\otimes \mathds{I}_\mc{D}) \rho_{\mc{D}\otimes\mc{Q}} (\ket{\downarrow}\otimes \mathds{I}_\mc{D})\\
&\quad+(\bra{\uparrow}\otimes \mathds{I}_\mc{D} )\rho_{\mc{D}\otimes\mc{Q}} (\ket{\uparrow}\otimes \mathds{I}_\mc{D})\,,
\end{split}
\end{equation}
and we immediately see that we do in fact have the correct choice of projection operators $P_1$ and $P_2$.

Next, in order to determine $B$ which defines the Kronecker product such that $\rho_{\mc{D}\otimes\mc{Q}} = \rho_\mc{D} \otimes \rho_\mc{Q}$ where $\rho_\mc{Q}=\sum_i p_i \ket{i}\bra{i}$ so we arrive at
\begin{equation}
\rho_{\mc{D}\otimes\mc{Q}} = \rho_\mc{D} \otimes \sum_i p_i  \ket{i}\bra{i} = \sum_i p_i (\mathds{I}_\mc{D} \otimes \ket{i} ) \rho_\mc{D} ( \mathds{I}_\mc{D} \otimes \bra{i})\,.
\end{equation}
In the case where our initial state is a pure state we further have
\begin{equation}
\rho = (\mathds{I}_\mc{D} \otimes \ket{i} ) \rho_\mc{D} ( \mathds{I}_d \otimes \bra{i})  = B \rho_\mc{D} B^{\dagger}\,,
\end{equation}
and we see we have identified the correct operator $B = \mathds{I}_\mc{D} \otimes \ket{i}$.

Overall, however, we are concerned with solutions in the dual-basis represented in Fig.~\ref{fig:Diagram2} spanned by the eigenstates $\ket{\psi_n}$ which can be expanded in terms of particle in a box eigenstates as
\begin{equation}
\begin{split}
&\ket{\psi_n}=\sum_i (a_{i,n}\ket{\phi_i} \otimes \ket{\downarrow} + b_{i,n}\ket{\phi_i}\otimes\ket{\uparrow})\\
&\quad \quad = \sum_i (a_{i,n}\ket{\phi_i^0} + b_{i,n}\ket{\phi_i^1})\,,
\end{split}
\end{equation}
where $a_{i,n}$ and $b_{i,n}$ are defined as
\begin{equation}
\begin{split}
& a_{i,n}=(\bra{\phi_i} \otimes \bra{\downarrow})\ket{\psi_n} = \int_{-L}^{0}dx \psi_n(x)\phi_i(x+L)\\
& b_{i,n}=(\bra{\phi_i} \otimes \bra{\uparrow})\ket{\psi_n} = \int_{0}^{L}dx \psi_n(x)\phi_i(x)\,.
\end{split}
\end{equation}

Now we are prepared to obtain operators which can be used in Eq.~(\ref{eq:CPTPMap}) along with our choice of basis.
Since the operators always show up in pairs, i.e. $BP_1$ or $BP_2$ we will forgo a derivation of each operator and for the sake of brevity only derive the form of $BP_1$ in the eigenbasis of the dual state if we have the initial preparations of qubits $\rho_n^1(0)$ \eqref{eq:prep}.
Using $BP_1 =(\mathds{I}  \otimes \ket{\downarrow})( \bra{\downarrow} \otimes \mathds{I}) $ and inserting a complete set of states, we have
\begin{equation}
\begin{split}
&(BP_1)_{lk} = \bra{\psi_l}BP_1\ket{\psi_k}\\
& = \sum_{ij} \left( a_{i,l}^* \bra{\phi_i}\otimes \bra{\downarrow}+b_{i,l}^* \bra{\phi_i}\otimes \bra{\uparrow}\right) \left(\mathds{I}  \otimes \ket{\downarrow} \right) \\
&\quad \quad \quad \times  \left( \bra{\downarrow} \otimes \mathds{I} \right) \left( a_{jk}\ket{\phi_j}\otimes \ket{\downarrow}+b_{jk}\ket{\phi_j}\otimes \ket{\uparrow} \right)\\
&= \sum_i a^*_{il}b_{ik}\,.
\end{split}
\end{equation}
All other operators needed to calculate the CPTP map in this model can be calculated in a similar manner.

\paragraph*{Finite dimensions and numerics.}

Finally, we conclude with a more technical remark: Strictly speaking the Hilbert space of the particle in a box is infinite dimensional . However, for explicit calculations we must restrict ourselves to a Hilbert space spanned by the lowest $M$ eigenvectors. Due to computational limitations all explicit calculations in the present work are done with $M=40$ eigenvectors. An additional complication that arises by this truncation is that, as defined, the CPTP map above is no longer trace preserving as the Hilbert space is no longer complete. To remedy this, we re-normalize the trace of the density operator between steps 2 and 3, i.e. $\rho \rightarrow \rho/\tr{\rho}$.

\section{Fixed Points}
\label{AppendixB}

\begin{figure}
	\includegraphics[width=.48\textwidth]{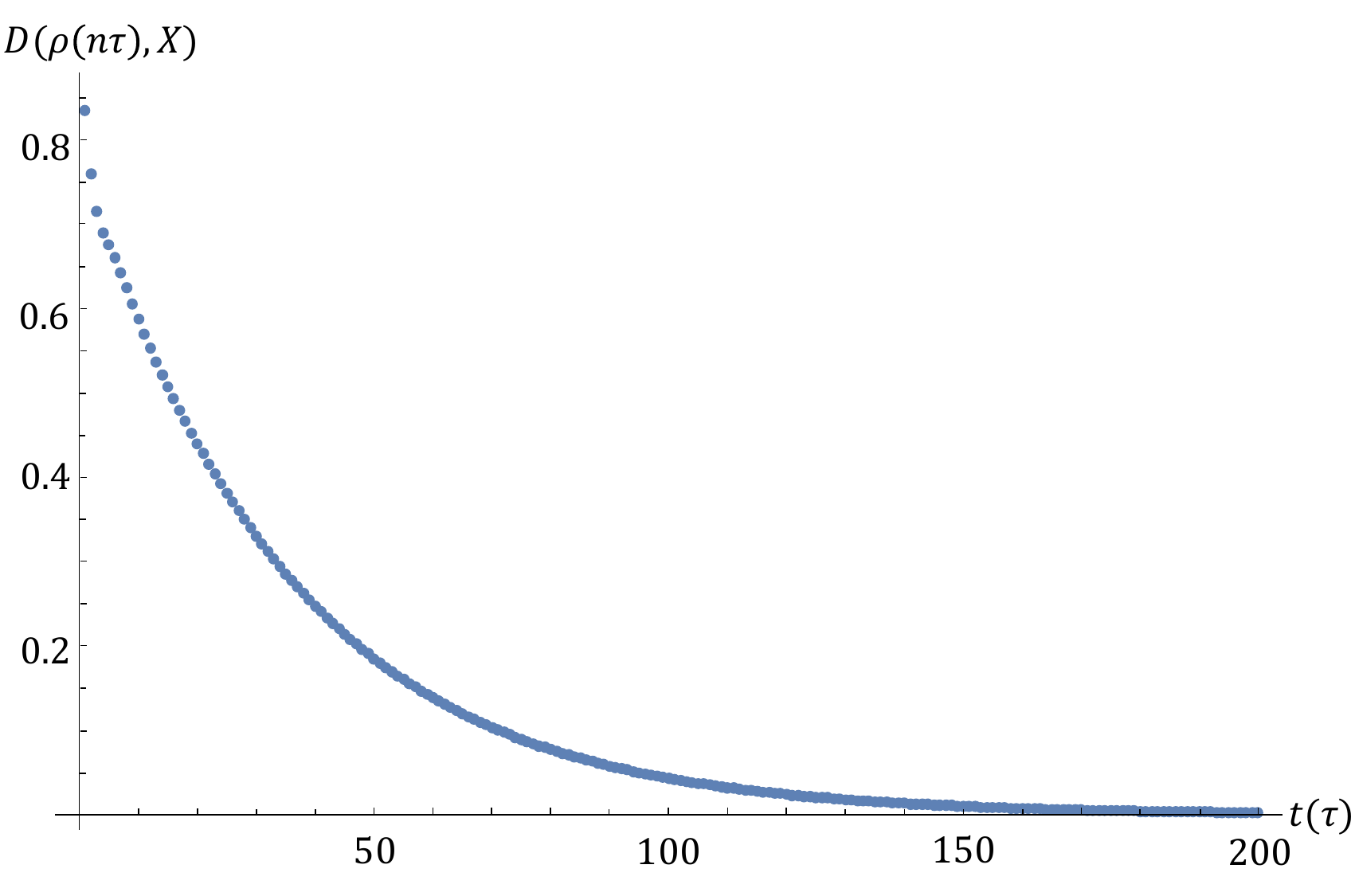}
	\caption{\label{Kdc}
	Trace distance between the demon state and the fixed point, $D(\rho(n \tau),X)$, as a function of repeated qubit interactions in time.  Here $M=10$  and $\tau_s = .94 \tau$.
}
\end{figure}

This appendix is dedicated to a numerical proof that the above constructed CPTP indeed relaxes towards its fixed point.To this end, we consider 
\begin{equation}
\begin{split}
 U(\tau) \left( \frac{B P_1 X P_1^\dagger B^\dagger +  B P_2 XP_2^\dagger B^\dagger }{\text{Tr}\left( B P_1 X P_1^\dagger B^\dagger +  B P_2 XP_2^\dagger B^\dagger \right)} \right) U^\dagger(\tau) 
 \\= X
 \end{split}
\end{equation}
where $X$ is the fixed point. Solving this equation requires us to solve a system of $M^2$ equations for $M^2$ unknowns and is done numerically. To understand the convergence we recursively apply our CPTP map to generate $\rho_\mc{D}(n\tau)$ and compare this to $X$ via the trace (Kolmogorov) distance \cite{Nielsen2010} given by
\begin{equation}
D(\rho(n \tau),X)= \frac{1}{2}\text{Tr}\left| \rho(n \tau) -  X\right|
\end{equation}
The result is illustrated in Fig.~\ref{Kdc}. We observe the trace distance goes to zero as $N$ becomes large and our system does indeed converge on the fixed point.  

\section{Zero current for classically mixed memories \label{app:zero}}

As a final consistency check we also computed the current for quantum memories that are prepared in classically, completely mixed states, i.e., it is equally likely to find the incoming qubit in either $\ket{\uparrow}$ or  $\ket{\downarrow}$. This situation is described by $\rho_n(0) = \rho_n^3(0)$. 
\begin{figure}
	\includegraphics[width=.48\textwidth]{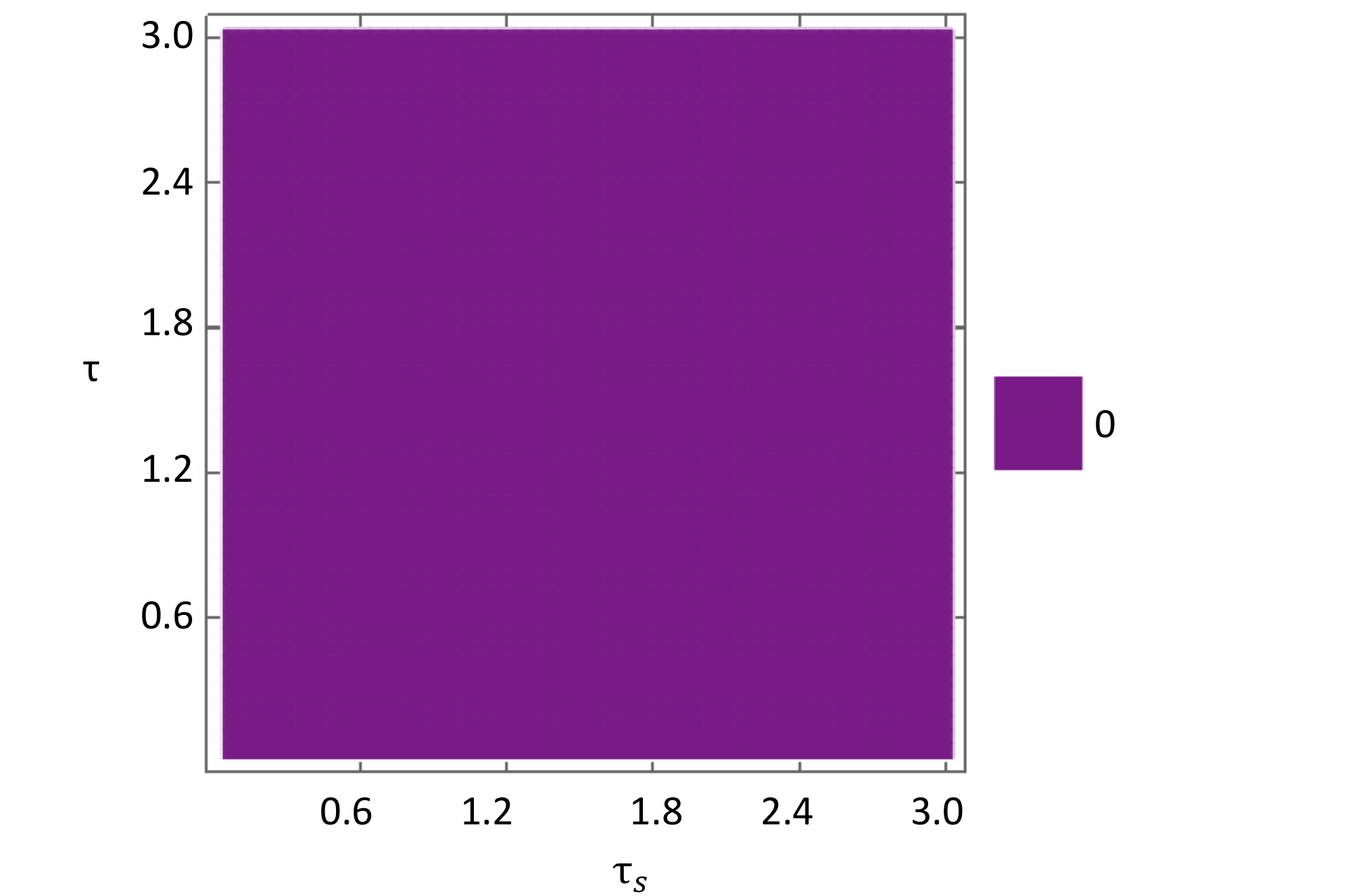}
	\caption{\label{p3}(color online) Total current $\bar{\Phi}$ for $\rho_n(0) = \rho_n^3(0)$ as a function of qubit interaction time $\tau$ and the characteristic time $\tau_s$ where both $\tau$ and $\tau_s$ are given in arbitrary units.  	
}
\end{figure}

As expected we found (numerically) that the resulting current is zero for any choice of parameters.

\bibliography{cite} 

\end{document}